\documentstyle[epsf,12pt,a4]{article}

\begin{document}

\title{ CMB Polarization During\\
the secondary Ionization of Matter}

\author{ M.V.Sazhin and A.V.Toporensky}
\date{
Sternberg Astronomical Institute, Universitetsky prospect,13, Moscow 119899,
Russia}

\maketitle

\date{ }
The cosmic microwave background (CMB) polarization, whose experimental
detection must become the next natural step in the study of the CMB
radiation, depends on the dynamics of hydrogen recombination in the
Universe, on the relation between the power of scalar perturbations
and gravitational waves, and on the presence or absence of secondary
ionization. The spectrum of the CMB anisotropy is currently known
with an accuracy that is high enough to draw some conclusions about
the cosmological parameters and about the presence of secondary
ionization of matter. The reduction of observational data strongly
suggests the presence of secondary ionization. We consider the effect
of this ionization at redshifts $z<100$ on the CMB polarization
generated by the scalar mode on various angular scales.

\section{Introduction}
\baselineskip=1.3\baselineskip
In recent years, measurements of the large- and intermediate-scale
anisotropy in the CMB radiation have reached an accuracy that allows
us to draw conclusions about the global parameters of the Universe and,
in particular, about the presence or absence of secondary recombination.
For example, de Bernardis {\it et al.}
 (1997) discussed the interpretation
of the observed spectrum of CMB fluctuations over a wide spectral range
(for $l \sim 2-600$). One of the possible conclusions of de Bernardis
{\it et al.}
 is that the model for the spectrum of primordial
perturbations must include secondary ionization. In this case, the
observes who will measure the CMB polarization will encounter a picture that
will differ from the standard scenario for the formation of
polarization.  We calculated the amplitude and angular dependence of
polarization measurements in the model with secondary ionizations.

The formation of CMB polarization in the standard model has been
studied extensively (Polnarev 1986; Bond {\it et al}.  1994; Harrari
 and Zaldarriaga 1993; Sarzin and Benitez 1995). In addition to the
standard models for the formation of the large-scale structure of the
Universe, models with unstable particles, one of whose decay channels
is photodecay with the release of photons that are hard enough to
ionize the primordial matter, have been considered by Berezhiani {\it
et al.} (1990), Sciama (1990), and Sakharov and Khlopov (1992). The
parameters of these particles, such as the half-life for the
photochannel and the energy of released photons, depend on a specific
physical theory and vary over a wide range.  In this case, the optical
depth may become significant, and the secondary formation of
polarization is possible.

\section{Analysis of the kinetic equation}

We assume the Friedmann model for the Universe with $\Omega_0=1$ and
$\Lambda=0$, $\Omega_{b}=0.05$ and postulate a dust-dominated equation
of state with $p=0$.

The relation for the interval with metric fluctuation is
$ds^2=a^2(\eta)(d
\eta^2-(\delta_{\alpha \beta} +h_{\alpha
 \beta})dx^{\alpha}dx^{\beta})$, where the small quantity
 $h_{\alpha \beta}$ satisfies the Einstein equation. Three independent
types of metric perturbation are recognized: scalar, vector and tensor.
In this paper we analyze the effect of only scalar perturbations or,
more precisely, of growing modes of adiabatic perturbations on the CMB
radiation.  The perturbation spectrum is considered in
Harrison-Zeldovich form with the spectral index $n=1$ (Starobinsky
1984).

Below, we use the standard notation for scalar correction to the metric
and their Fourier components:

$$
h_{\alpha \beta}(k, \eta)=h(k)(k\eta)^2\gamma_{\alpha}\gamma_{\beta},
$$
where $h(k)$ are the stochastic variations that show a
$\delta$ correlation with the power spectrum
$$
\langle h(k)h^{*}(q)\rangle=\frac{P(k)}{k^3}\delta (k-q),
$$
Here,
$P(k)=P_0\,k^{n-1}$,

and $n$  is the spectral index.

In order to calculate the degree of polarization, we solve the kinetic
equation for the symbolic vector

$$
\delta=\pmatrix{\delta_l\cr \delta_r\cr \delta_u}.
$$
The form of the kinetic equation and its details can be found in Basko
and Polnarev (1980) and in the book of Chandrasekhar (1960).
As it was shown by Polnarev (1986), Sazhin and Benitez (1995),
Harrari and Zaldarriaga (1993), and Crittenden {\it et al.} (1993),
the solution of the kinetic equation for the function
 $\delta$  in the case of plane waves will suffice to determine the anisotropy
and polarization of the CMB radiation for an arbitrary perturbation
spectrum.  The equation in Fourier components for one of the waves with
the wave vector $k$ is

\begin{equation}
\frac{\partial \delta}{\partial \eta} +e^{\alpha}\frac{\partial
\delta}{\partial x^{\alpha}} =\frac{1}{2}\frac{\partial h_{\alpha
\beta}}{\partial \eta}e^{\alpha}e^{\beta} -a(\eta)N_e \sigma (\delta
-\int P(\Omega, \Omega') \delta(\Omega)d\Omega).
\end{equation}

We choose a coordinate system in which the perturbations propagate
along the $z$ axis. The angle between the axis of photon propagation
and the $z$ axis is denoted by $\theta$.
We than have the following relation for the driving force on the
right side of the kinetic equation which describes the formation
of polarization (Basko and Polnarev 1980, Polnarev 1986):

\begin{eqnarray}
F=h k^2\eta (\mu^2-1/3),
\end{eqnarray}

where $h$ is the amplitude of the scalar perturbations, and $\mu
 =\cos(\theta)$.

We eliminated the monopole anisotropy component to obtain the second
Legendre polynomial. For convenience, we choose the following form for
$\delta$:

\begin{eqnarray}
\delta=\alpha \cdot (\mu ^2-1/3) \pmatrix{1\cr 1\cr 0} +
\beta \pmatrix{1\cr -1\cr 0}(1-\mu ^2)
\end{eqnarray}

After substituting this vector into the kinetic equation for
$\delta$ (see, e.g., Basko and Polnarev 1980; Polnarev 1986), we
obtain a system of integrodifferential equations. The solutions of this
system are analyzed, in particular, by Gibilisco (1995,1996) and Seljak
and Zaldarriaga (1996) by numerical and analytical methods.  Since the
use of numerical methods always raises the question of generality of
the solutions, we use the analytical method of expansion in terms of a
small parameter -- the optical depth $\tau$. This expansion yields
estimates of the effect which are general enough and essentially
independent of a particular scenario. The scenario with an optical
depth that is small compared to unity produced by the secondary
ionization is currently most popular (de Bernardis {\it et al.}
(1997).The case of a large optical depth produced dy the secondary
ionization, where the approximation of instantaneous recombination can
be used to obtain an analytical solution was considered by us in our
previous paper (Sazhin and Toporensky 1995).

Substituting (3) into the kinetic equation, we obtain the system of equations

\begin{eqnarray}
\frac{d\alpha}{d\eta}+ik\mu\alpha=F(k,\eta)-g(\eta)[\alpha-\frac{9}{16}I(k,\eta)],\\
\frac{d\beta}{d\eta}+ik\mu\beta=-g(\eta)[\beta+\frac{9}{16}I(k,\eta)],\\
 I(k,\eta)=\int^1_{-1} d\mu [\alpha(k,\eta,\mu)(\mu^2-{1\over 2})^2-
\beta(k,\eta,\mu)(1-\mu^2)^2] .
\end{eqnarray}

We solve this system by expanding it in terms of $g(\eta)$.
In the zeroth approximation (i.e., in the approximation in which the
quantities $\alpha, \beta$ correspond to the anisotropy and
polarization produced during the primary recombination) $\alpha$ is
considerably larger in absolute value than $\beta$. As a result, when
considering the first approximation, the effect of $\beta$ on $\alpha$
is considerably smaller than the inverse effect. The correction to $\beta$
that follows from the first term in the expansion of our system may
exceed significantly (for a given value of the wave vector) the
zero-approximation value for $\beta$.

In the zeroth approximation for $g(\eta)$, the solution for $\alpha$ is
\begin{equation}
\alpha_0=\exp (-ik\mu\eta) \int F(k,\hat \eta, \mu) \exp (ik\mu
\hat \eta)d\hat \eta .
\end{equation}
The corresponding solution for $\beta$, as follows from the discussion above,
can be taken in the form $\beta_0=0$.

The solution for $\beta$ in the first approximation is now
\begin{equation}
\beta_1(k,\eta)=-\frac{9}{16}\exp (-ik\mu\eta) \int^{\eta}_{\eta_r}
g(\eta)I_0(k,\hat \eta) \exp (ik\mu \hat \eta)d\hat \eta ,
\end{equation}
where

\begin{equation}
I_0(k,\eta)=\int^1_{-1} d\mu [\alpha(k,\eta,\mu)(\mu^2-{1\over 2})^2]
\end{equation}

\section{Secondary ionization and the degree of polarization}

Let us consider the model of instantaneous secondary ionization
in which the degree of ionization is equal to zero for redshifts
$z>z_{sr}$ and unity for $z<z_{sr}$.
In this case, the function $g(\eta)$ is described by a relation of the form
\begin{equation}
g(\eta)=\sigma_T  n_{sr} (\frac{\eta_{sr}}{\eta_0})^6
(\frac{\eta_0}{\eta})^4
\end{equation}
Introducing the baryon density $\Omega_b$, we can write the preceding
equation in the form
 \begin{equation}
 g(\eta)=\frac{3H^2}{8\pi G}
\frac{\sigma_T}{m_p} \Omega_b (\frac{\eta_0}{\eta})^4
 \end{equation}

Denoting $k (\eta-\eta_r)$ by $x$ and taking the integral in (9),
we obtain

$$
I_0=h_A (\frac{8}{45}-f(x)-k \eta_r \frac{ \partial f(x)}{ \partial
x}) ,
$$
 where the function $f(x)$ is given by

$$
f(x)=\sqrt(2 \pi) [8 \frac{J_{5/2}(x)}{x^{5/2}} - \frac{8}{3}
\frac{J_{3/2}(x)}{x^{3/2}} + \frac{4}{9} \frac{J_{1/2}(x)}{x^{1/2}}].
$$
Thus, the degree of polarization $\beta$ is the product of the integral
over $\eta$ and three constants ($H,\Omega_b$ ¨ $h_A$) that depend on the
cosmological scenario:

$$
\beta(k, \mu, \eta)=-{9 \over 16} exp(-ik\mu \eta)\frac{3H^2}{8\pi G}
\frac{\sigma_T}{m_p} \Omega_b h_A \int_{\eta_r}^{\eta_0}d\eta \;
(\frac{\eta_0}{\eta})^4 f(x) exp(ik\mu \eta).
$$

Let us introduce a polarized anisotropy component given by
 $$
p={\delta_l-\delta_r \over 2} T
$$

A plot of this quantity against the wave vector $k$
is shown in the figure. Note that the half-width at half-maximum
(HWHM) of the spectral curve is essentially independent of the
problem parameters and is determined only by the position of the
maximum $k_{max}$ of the curve; it is equal to $\Delta k \approx 0.63 k_{max}$.

\begin{figure}
\epsfxsize=\hsize
\centerline{\epsfbox{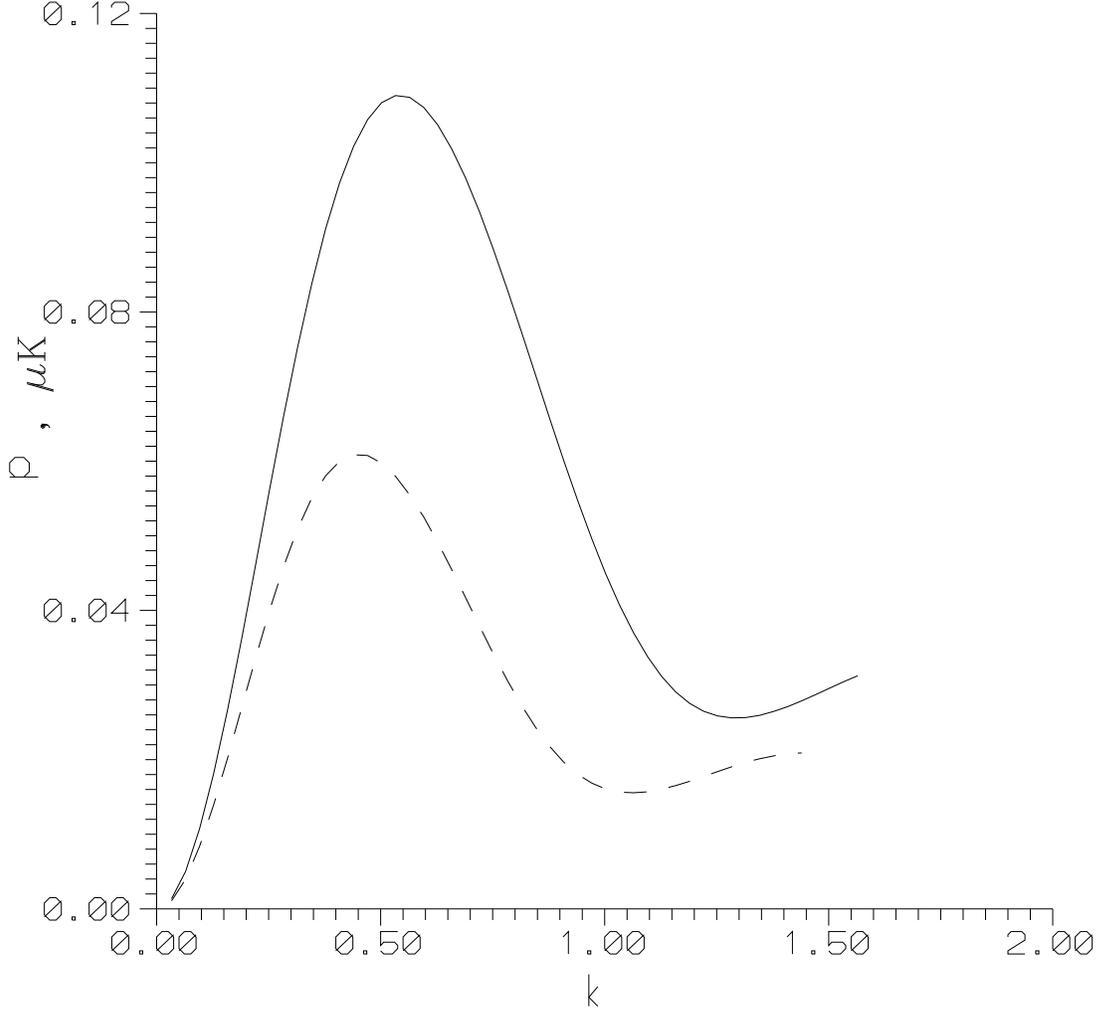}}
\caption{The quantity $p$ versus the wave vector of scalar
perturbations for $\eta_{sr}=6$ (solid line) and $\eta_{sr}=7$
(dashed line).}
\end{figure}

Let us also define the angular scale in terms of $k_{max}$  as follows:
$$
\theta_{max}=2^{\circ} \frac{2\pi}{k_{max}}.
$$
It should be emphasized that this angular scale is not optimal for
choosing the parameters of the antenna that measured the polarization.
This quantity characterizes the scale of fluctuations of the
polarized component in the sky. To choose an optimum angular scale
requires optimization with allowance for the antenna beam and the
method of measurements.

The data in the table were obtained for
$\Omega_b=0.05, H=50$km/(s Mpc) and $h_A$ normalized to the COBE
results (Smoot 1997). The first and second columns give the conformal
time, which corresponds to the time of secondary ionization and the
redshifts, respectively. The third column contains the values for the
quantity $p_{max}$ defined as follows:
$$
 p_{max} =\sqrt{|\beta(k_{max})|^2} T  (\mu K) $$

The values of $\theta_{max}$ are given in the fourth column.
Finally, the last column lists the rms values of the polarization
in $\mu K$ as estimated from the relation
$\delta T_p= {1 \over
2}\beta_{max} \Delta k$.

\bigskip
\begin{center}
\begin{tabular}{|c|c|c|c|c|}
\hline
$\eta_{sr}$ & $z_{sr}$ & $p_{max}\; (\mu K)$ & $\theta_{max}$ &
$\delta T_P\; (\mu K)$\\
\hline
\hline
3.0 & 100 & 1.62 & $9^{\circ}.6$ & 0.70\\
\hline
3.5 & 72 & 0.85 & $12^{\circ}.4$ & 0.30\\
\hline
4.0 & 55 & 0.49 & $15^{\circ}.2$ & 0.14\\
\hline
5.0 & 35 & 0.20 & $21^{\circ}.6$ & 0.041\\
\hline
6.0 & 24 & 0.10 & $28^{\circ}.5$ & 0.016\\
\hline
7.0 & 17 & 0.05 & $36^{\circ}.1$ & 0.007\\
\hline
\end{tabular}
\end{center}
\vskip3mm

The existence of secondary ionization results in a change in the angular
dependence of the amplitude of the polarized component. In particular,
an additional peak appears on intermediate angular scales, which are
most convenient, for example, from the standpoint of the SPORT
experiment (Cortiglioni {\it et al.} 1997) and several other currently
functioning or planned experiments.

\lineskip2cm

\begin{center}
\bf References
\end{center}
\begin{itemize}
\item[] Basko, M.M.and Polnarev, A.G. // MNRAS,
1980,  V.{\bf 191}, P. 47.

\item[] Berezhiani, Z.G., Khlopov, M.Yu. and Khomeriki, R.R. // Sov.J.Nucl.Phys
1990,  V.{\bf 52}, P.65.

\item [] de Bernardis P., Balbi A., De Gasperis G. {\it et al.}
 // Astrophys. J., 1997,V.{\bf 480}, P.1.

\item [] Bond J.R., Davis R.L., Steinhard P.L. //
Phys.Rev.Lett.,1994,V.{\bf 72}, P.13.

\item[] Chandrasekhar, S. // Radiative Transfer, New York: Dover, 1960.
\item[] Cortiglioni S., Sironi G., Strukov I.A. {\it et al.}  //
  Sky Polarization Observatory (SPORT),
TESRE, 1997.

\item[] Crittenden R., Davies R.L., Steinhard P.J. //
      Astrophys.J. (Letters), 1993, V.{\bf 417}, L13.

\item[] Gibilisco M. // Intern. J. Modern
Phys. A., 1995, V.{\bf 10}, P.3605.

\item[] Gibilisco M. // Astrophys. Space Sci., 1996,
V.{\bf 235}, P.75.

\item[] Harrari D.D., Zaldarriaga M. //
Phys.Lett., 1993,  V.{\bf 315B}, P.96.

\item[] Polnarev, A.G. //  Sov. Astron., 1986, V.{\bf 29}, P.607.

\item[] Sakharov, A.S. and Khlopov, M.Yu. // Sov. J. Nucl. Phys. 1992,
 V.{\bf 55},P.1063.

\item[] Sazhin, M.V., Benitez N. // Astron. Lett. and
Communications, 1995, V.{\bf 32}, P.105.

\item[] Sazhin, M.V. and Toporensky, A.V. // Astron. Lett., 1995,
V.{\bf 21}, P.498.

\item[] Sciama D.W. // MNRAS, 1990, V.{\bf
244}. P.1P

\item[] Seljak U., Zaldarriaga M. //
astro-ph/9609169

\item[] Smoot G.F. //   Microwave Background Anisotropies,(ed. S.
F.R.Bouchet, R.Gispert, B.Guiderdoni, J. Tran Thanh Van),
 Gif - Sur Yvette Cedex France, Editions Frontieres,1997, P.3.

\item[] Starobinsky, A.A. // Sov. Astron. Lett. 1984, V.{\bf 9},
P.302.

 \end{itemize}

\end {document}